\documentclass[11pt]{article}
\usepackage{jheppub}

\newcommand\fft[2]{{\frac{#1}{#2}}}
\newcommand\ft[2]{{\textstyle\frac{#1}{#2}}}
\newcommand\nn{\nonumber}

\begin{document}

\preprint{MCTP-13-28}

\title{\boldmath $1/N^{2}$ corrections to the holographic Weyl anomaly}

\author[a]{Arash Arabi Ardehali,}
\author[a]{James T. Liu,}
\author[b]{and Phillip Szepietowski}

\affiliation[a]{Michigan Center for Theoretical Physics, Randall Laboratory of Physics,\\
The University of Michigan, Ann Arbor, MI 48109--1040, USA}
\affiliation[b]{Department of Physics, University of Virginia,\\
Box 400714, Charlottesville, VA 22904, USA}

\emailAdd{ardehali@umich.edu}
\emailAdd{jimliu@umich.edu}
\emailAdd{pgs8b@virginia.edu}

\abstract{We compute the $\mathcal O(1)$ contribution to holographic
$c-a$ for IIB supergravity on AdS$_5\times S^5/\mathbb Z_n$ and on
AdS$_5\times T^{1,1}/\mathbb Z_n$.  In both cases, we find agreement
with the dual field theory results, thus providing $1/N^2$ checks of
AdS/CFT with reduced supersymmetry. Since the holographic
computation involves a sum over shortened multiplets in the KK
tower, we provide some details on the $S^5$ and $T^{1,1}$ spectra in
a form that is convenient when considering their $\mathbb Z_n$
orbifolds. The computation for the even $\mathbb Z_n$ orbifolds of
$S^5$ includes a sum over the multiplets in the twisted sector that
is essential for obtaining agreement with the dual field theory.}

\maketitle
\flushbottom

\section{Introduction}

The AdS/CFT correspondence relates, among other things, strongly
coupled superconformal gauge theories to their dual string theories
in weakly curved AdS backgrounds. Thus, in principle, one requires
some knowledge of the strongly interacting field theory to test the
duality.  While this is often a challenging situation in general, in
some cases it is possible to obtain exact results at strong
coupling.  These cases include the study of protected operators and
BPS states as well as anomalies.

On the CFT side of the duality, the theory may be characterized
by two central charges $a$ and $c$ in four dimensions.  Moreover,
data on these central charges can be obtained at strong coupling
based on 't~Hooft anomaly matching and supersymmetry
arguments \cite{Anselmi:1997am}.  Assuming the CFT admits an AdS dual,
it is then possible to reproduce the central charges through the
holographic Weyl anomaly \cite{Henningson:1998gx}.  At large $N$, where the dual
string theory can be well approximated by classical supergravity,
the result $c=a=\mathcal O(N^2)$ exactly matches the CFT
result at leading order.

Here we are interested in going beyond the leading order in
comparing the central charges on both sides of the duality.  In
particular, we focus on four-dimensional quiver gauge theories
dual to string theory on AdS$_5\times S^5/\mathbb Z_n$ as
well as on AdS$_5\times T^{1,1}/\mathbb Z_n$.  For the $\mathbb Z_n$
orbifold of $S^5$, a gauge theory computation gives
\begin{eqnarray}
\begin{cases}
c=a=\fft{N^2-1}4,& n=1;\\
c=\fft{N^2}2-\fft13,\kern4em a=\fft{N^2}2-\fft5{12},& n=2;\\
c=n\left(\fft{N^2}4-\fft18\right),\qquad a=n\left(\fft{N^2}4-\fft3{16}\right),& n\ge3,
\end{cases}
\label{eq:c,aS5Zn}
\end{eqnarray}
which matches the leading order holographic Weyl anomaly computation
\begin{equation}
c=a=\fft{N^2}4\fft{\pi^3}{\mathrm{vol}(S^5/\mathbb Z_n)}=n\fft{N^2}4.
\end{equation}
The corresponding expressions for $T^{1,1}/\mathbb Z_n$ are
\begin{equation}
c=n\left(\fft{27}{64}N^2-\fft14\right),\qquad a=n\left(\fft{27}{64}N^2-\fft38\right),
\label{eq:c,aT11Zn}
\end{equation}
which again matches the holographic computation at leading order.

In order to reproduce the $\mathcal{O}(1)$ terms
($\mathcal{O}(1/N^{2})$ relative to the leading $\mathcal{O}(N^{2})$
terms) holographically, we must go beyond the tree level and
consider loop corrections to the bulk effective action. As argued in
\cite{Liu:2010gz}, these loop effects fall into two categories: $i$) massive
string states running in the loop, and $ii$) massless ten-dimensional supergravity
states in the loop. From a five-dimensional point of view, the
latter includes not just the supergravity states, but also those
from the Kaluza-Klein tower obtained from compactifying IIB
supergravity on either $S^5/\mathbb Z_n$ or $T^{1,1}/\mathbb Z_n$.

The case of $\mathcal{N}=4$ SYM has been investigated in
\cite{Bilal:1999ph,Bilal:1999ty,Mansfield:2000zw,Mansfield:2002pa,Mansfield:2003gs},
where it was shown that the shift $N^2\to N^2-1$ can be accounted
for by considering the one-loop contribution from the Kaluza-Klein
tower on $S^5$.  In the approach of \cite{Mansfield:2000zw,Mansfield:2002pa,Mansfield:2003gs},
the subleading holographic Weyl anomaly $\delta\mathcal{A}$ may
be computed from the expression
\begin{equation}
\delta\mathcal{A}=-\sum(-1)^{F}\frac{(E_0-2)a_{2}}{32\pi^{2}},
\label{eq:MNU}
\end{equation}
where the sum is over all the bulk KK states that could run in the
loop. Here $a_{2}$ are four dimensional heat-kernel coefficients for
the transverse space in the bulk (which has the same geometry as the
regularized boundary), and $E_0$ is the lowest energy eigenvalue
labeling the AdS representation of the field (corresponding to the
conformal dimension $\Delta$ in the CFT dual). Using the appropriate
heat-kernel coefficients, both $c$ and $a$ can be independently
extracted from (\ref{eq:MNU}), provided the full KK spectrum of the
five-dimensional bulk theory is available. In particular, when
applied to the spectrum of IIB supergravity on AdS$_5\times S^5$
\cite{Gunaydin:1984fk,Kim:1985ez}, this expression successfully
reproduces both $c$ and $a$ of $\mathcal N=4$ SYM beyond the leading
order \cite{Mansfield:2002pa}.

There are, however, several practical issues in applying
(\ref{eq:MNU}). Firstly, since the sum is over an infinite number of
states in the KK tower, it needs to be regulated in some manner.
Although physical quantities should not depend on choice of
regulator, it is not entirely clear how this would work out in
general.  We will have more to say about this below. Secondly, the
full linearized KK spectrum may be difficult to obtain for
compactifications with reduced supersymmetry. This is not an issue
for orbifolds of $S^{5}$ and $T^{1,1}$ considered here, whose
complete linearized KK spectrum can be found from the known spectrum
on their covers. However it becomes a difficulty when considering
other interesting compactification spaces, such as $Y^{p,q}$ or $L^{a,b,c}$
manifolds \cite{Gauntlett:2004yd,Cvetic:2005ft}.

For the case of $\mathcal N=2$ supergravity in AdS$_5$ (dual to
$\mathcal N=1$ superconformal field theory), the KK tower organizes
itself into a combination of long and shortened multiplets.  In
Ref.~\cite{Ardehali:2013gra}, we demonstrated that while long multiplets
will contribute to the individual central charges $c$ and $a$
obtained from (\ref{eq:MNU}), their contribution vanishes for the
difference $c-a$. In particular, this allows us to compute the
difference using only knowledge of the protected data in the bulk.
Note that, from the field theory point of view, the combination $c-a$ shows up in
both the Weyl anomaly and the $R$-current anomaly via
\cite{Anselmi:1997am,Anselmi:1998zb}
\begin{eqnarray}
\langle T^{\mu}_{\mu}\rangle&=&\frac{c-a}{16\pi^{2}}R_{\mu\nu\rho\sigma}^{2}+\cdots,\nn\\
\langle\partial_{\mu}\sqrt{g}R^{\mu}\rangle&=&\frac{c-a}{48\pi^{2}}
\epsilon^{\mu\nu}{}_{\rho\sigma}R_{\mu\nu\alpha\beta}R^{\rho\sigma\alpha\beta}+\cdots.
\end{eqnarray}
This relation of $c-a$ to the $R$-current anomaly seems to be
connected to the vanishing contribution from long multiplets. Note
also that the combination $c-a$ gives the full Weyl anomaly if the
field theory lives on a Ricci-flat manifold.

Since $c=a$ at leading order, we can rewrite (\ref{eq:MNU}) in the
useful form
\begin{equation}
c-a=-\fft12\sum(-1)^F(E_0-2)a_2\Big|_{R_{\mu\nu\rho\sigma}^2\mathrm{~term}},
\label{eq:c-aexpr}
\end{equation}
where the sum is only over states in shortened representations in
the KK tower.  Here it is worth noting that the coefficient of the Riemann-squared
term in $a_2$ has a particularly simple form \cite{Christensen:1978gi}.  For a
four-dimensional field transforming in the irreducible representation $(A,B)$ of
the Lorentz group, the expression is
\begin{equation}
180a_2\Big|_{R_{\mu\nu\rho\sigma}^2\mathrm{~term}}
=d(A,B)\left(1+f(A)+f(B)\right),
\end{equation}
where $d(A,B)=d(A)d(B)=(2A+1)(2B+1)$ is the dimension of the representation
and $f(X)=X(X+1)(6X(X+1)-7)$.  The crucial observation is that the expression
for $c-a$ then splits into a sum of factorized pieces
\begin{equation}
c-a=-\fft1{360}\sum(E_0-2)(-1)^{2A+2B}\left[d(A)d(B)+(d(A)f(A))d(B)+d(A)(d(B)f(B))\right].
\label{eq:ChristensenDuff}
\end{equation}
The reason this vanishes for long multiplets is that such multiplets carry an equal number
of integer and half-integer helicity states labeled by $A$ and $B$, so that both
$\sum(-1)^{2A}d(A)=0$ and $\sum(-1)^{2B}d(B)=0$, along with $\sum E_0(-1)^{2A}d(A)=0$
and $\sum E_0(-1)^{2B}d(B)=0$.  (See e.g. \cite{Freedman:1999gp} for a summary of
unitary representations of $\mathrm{SU}(2,2|1)$ and shortening conditions.)
Note that this cancellation is independent of the explicit form of $f(X)$.

Shortened representations, however, do not have both $A$ and $B$ sums
vanishing. Chiral and semi-long II multiplets have
$\sum(-1)^{2A}d(A)=0$ and $\sum E_0(-1)^{2A}d(A)=0$, while
anti-chiral and semi-long I multiplets have the corresponding sums
over $B$ vanishing.  Since this is insufficient to ensure the
vanishing of all terms in (\ref{eq:ChristensenDuff}), $c-a$ will
receive a contribution from all short multiplets in the KK spectrum.
(In this case, the explicit form of $f(X)$ does enter when computing
the anomaly.)  This suggests that $c-a$ may be related to some sort
of indices on
either side of the duality%
\footnote{A similar suspicion was stated in \cite{Gadde:2011} that
the anomaly coefficients might be related to the superconformal
index on $S^{3}\times \mathbb{R}$. Since the anomaly coefficients
are sensitive to the detailed spectrum only at subleading order, we
consider it more likely that if such a relation exists, it would
relate the superconformal index to the \emph{subleading} part of
the anomaly coefficients and perhaps directly to $c-a.$}.

Focusing on the difference $c-a$, we see from (\ref{eq:c,aS5Zn})
that for $S^5/\mathbb Z_n$ orbifolds, the field theory gives
\begin{equation}
c-a=\begin{cases}0,&n=1;\\
\fft1{12},&n=2;\\
\fft{n}{16},&n\ge3.
\end{cases}
\label{eq:ftresult}
\end{equation}
The $n=1$ case corresponds to the round five-sphere, or equivalently
$\mathcal N=4$ SYM, and the result $c=a$ was reproduced on the
gravity side in \cite{Mansfield:2002pa} by regulating the sum
(\ref{eq:c-aexpr}). We studied the $S^5/\mathbb Z_3$ case,
corresponding to $\mathcal N=1$ $\mathrm{SU}(N)^3$ gauge theory, in
\cite{Ardehali:2013gra} and similarly found exact matching with $c-a=3/16$
on both sides of the duality.  In this paper, we extend our previous
result for $S^5/\mathbb Z_3$ to arbitrary $\mathbb Z_n$ orbifolds of $S^5$ and
again find exact matching with $c-a=n/16$.  For even orbifolds, a
contribution from the twisted sector is expected; this may be computed
by starting with the low energy effective description of the twisted sector
in terms of a $(2, 0)$ tensor theory in six dimensions, KK reducing
it to five dimensions and then applying Eq.~(\ref{eq:c-aexpr}) to the resulting
five-dimensional spectrum.

One issue that we have only alluded to so far is the contribution to the
bulk effective action from massive string states running in the loop.
As argued in \cite{Liu:2010gz}, such holographic contributions would show up
through higher-derivative corrections in the five-dimensional effective
action, and they should be added to (\ref{eq:c-aexpr}) to obtain the
complete subleading shift to $c-a$.  However, it turns out that these
massive string loop contributions vanish for compactifications on
$S^5$ and its $\mathbb Z_n$ orbifolds.  Therefore the exact matching
$c-a=n/16$ (or $c-a=1/12$ for $n=2$) is unaffected by massive string loop
considerations.

The issue of massive string loop corrections will arise for other
Sasaki-Einstein compactifications of IIB string theory.  In
particular, the computation in \cite{Liu:2010gz} predicts that such string
loops would contribute $1/24$ to $c-a$ for the conifold theory. In
order to investigate this possible contribution, in this paper we
also examine orbifolds of $T^{1,1}$.  Curiously, we find that the
sum of the KK tower in (\ref{eq:c-aexpr}) completely reproduces the
field theory result, so that massive string loop contributions are
in fact \emph{not} necessary (and so would ruin the matching if
included).  This presents a puzzle for the fate of the massive
string loop corrections.

This paper is organized as follows.  In the next section, we
consider IIB supergravity on AdS$_5\times S^5/\mathbb Z_n$ and find
the result $c-a=n/16$ (or $c-a=1/12$ for $n=2$), in agreement with
the field theory side of the duality.  We also elaborate on the
twisted states appearing in the cases with even $n$ and demonstrate
that they are necessary for the matching to work. In
section~\ref{sec:T11}, we examine IIB supergravity on AdS$_5\times
T^{1,1}/\mathbb Z_n$ and find $c-a=n/8$, which matches the field
theory result provided there are no further contributions from
massive string loops.  Finally, we conclude in section~\ref{sec:dis}
with some open questions.  Some details on the twisted sector of the
$S^5/\mathbb Z_2$ orbifold are presented in
Appendix~\ref{app:twisted}.

\section{Orbifolds of $S^{5}$}
\label{sec:S5}

Perhaps the best studied framework for AdS/CFT involves the duality between IIB
string theory on AdS$_5\times S^5$ and $\mathcal N=4$ super-Yang-Mills theory.
This system preserves 32 real supercharges, and the appropriate supergroup is
$\mathrm{SU}(2,2|4)$.  Application of (\ref{eq:MNU}) demonstrates that
the leading order Weyl anomaly $c=a=N^2/4$ gets shifted to $c=a=(N^2-1)/4$
\cite{Mansfield:2002pa}.  However, the difference $c-a$ continues to vanish because
of maximal supersymmetry.

Starting with AdS$_5\times S^5$, it is straightforward to consider the family of orbifold
models AdS$_5\times S^5/\mathbb Z_n$ that preserve a reduced amount of supersymmetry.
Here the orbifold $S^5/\mathbb Z_n$ is obtained by starting with $\mathbb C^3$ intersected
with the unit sphere and modding out by the $\mathbb Z_n$ action generated by
\begin{equation}
\Omega=\begin{pmatrix} \omega\cr &\omega\cr &&\omega^{-2}\end{pmatrix},
\label{eq:Znact}
\end{equation}
where $\omega^{n}=1$.  Since this element is contained in SU(3), the
orbifold generically preserves $\mathcal N=1$ supersymmetry in four
dimensions. Note, however, that for $n=2$ this element is in the
center of SU(2), so the $S^5/\mathbb Z_2$ orbifold actually
preserves $\mathcal N=2$ supersymmetry in four dimensions.  (The
$n=3$ case is also somewhat special, as the element is then in the
center of SU(3), a fact that we found useful in the analysis of
\cite{Ardehali:2013gra}).

\subsection{The spectrum and shortenings}

For orbifolds of $S^5$, the natural starting point is simply the
spectrum of IIB supergravity on the round $S^5$, originally obtained
in \cite{Gunaydin:1984fk,Kim:1985ez}.  Since here we are interested
in $\mathcal N=2$ supergravity in five dimensions, we rewrite the
$\mathcal N=8$ spectrum in $\mathcal N=2$ language that will be
convenient for further applications. This is shown in
Table~\ref{tbl:SU3KK}, where $\mathcal D(E_0,s_1,s_2;r)$ label the
irreducible representations of the superalgebra SU(2,2$|$1).

\begin{table}[t]
\centering
\begin{tabular}{|l|l|l|}
\hline
Supermultiplet&Representation&KK level\\
\hline
Graviton&$\sum_{k=0}^{p-2}\mathcal D(p+1,\ft12,\ft12;\ft13(2p-4k-4))(k,p-k-2)$&$p\ge2$\\
Gravitino I and III&$\sum_{k=0}^{p-1}\mathcal D(p+\ft12,\ft12,0;\ft13(2p-4k+1))(k,p-k-1)$&
$p\ge2$\\
&$+\sum_{k=0}^{p-1}\mathcal D(p+\ft12,0,\ft12;\ft13(2p-4k-5))(k,p-k-1)$&\\
Gravitino II and IV&$\sum_{k=0}^{p-3}\mathcal D(p+\ft32,\ft12,0;\ft13(2p-4k-9))(k,p-k-3)$&
$p\ge3$\\
&$+\sum_{k=0}^{p-3}\mathcal D(p+\ft32,0,\ft12;\ft13(2p-4k-3))(k,p-k-3)$&\\
Vector I&$\sum_{k=0}^p\mathcal D(p,0,0;\ft13(2p-4k))(k,p-k)$&$p\ge2$\\
Vector II&$\sum_{k=0}^{p-4}\mathcal D(p+2,0,0;\ft13(2p-4k-8))(k,p-k-4)$&$p\ge4$\\
Vector III and IV&$\sum_{k=0}^{p-2}\mathcal D(p+1,0,0;\ft13(2p-4k-10))(k,p-k-2)$&$p\ge2$\\
&$+\sum_{k=0}^{p-2}\mathcal D(p+1,0,0;\ft13(2p-4k+2))(k,p-k-2)$&\\
\hline
\end{tabular}
\caption{\label{tbl:SU3KK} The spectrum of IIB supergravity on $S^5$
written in terms of $\mathcal N=2$ multiplets, and with the
decomposition $\mathrm{SU}(4)\supset\mathrm{SU}(3)\times\mathrm
U(1)_r$.  The supermultiplets are given in the conventional notation
$\mathcal D(E_0,s_1,s_2;r)$ with the $\mathrm{SU}(3)$ representation
given in terms of Dynkin labels $(l_1,l_2)$ appended.}
\end{table}

For the holographic computation of $c-a$, however, only the
shortened spectrum of the theory is needed.  There are three
multiplet-shortening conditions, corresponding to conserved, chiral
(anti-chiral) and semi-long I (semi-long II) multiplets. Since these
conditions constrain the relation between $E_0$ and $r$, for a given
KK level $p$, only terms at the ends of the sums over $k$ in
Table~\ref{tbl:SU3KK} correspond to shortened states.  The shortened
spectrum is shown in Table~\ref{tbl:KKshort}.  In this table, we
also present the contribution of each short multiplet to $c-a$ as
obtained in \cite{Ardehali:2013gra}. As a check, we have summed over all
states shown in Table~\ref{tbl:KKshort}, and found a vanishing
correction to $c-a$, in agreement with the result of
\cite{Mansfield:2002pa} for the round $S^5$ (dual to $\mathcal N=4$
SYM).

\begin{table}[t]
\centering
\begin{tabular}{|l|l|l|l|l|}
\hline
Multiplet&KK&Shortened representation&Shortening&$c-a$ for one\\
&level&&type&\\
\hline
Graviton&$p=2$&$\mathcal D(3,\ft12,\ft12;0)(0,0)$&conserved&$-\ft58$\\
&$p>2$&$\mathcal D(p+1,\ft12,\ft12;-\ft23(p-2))(p-2,0)$&SLI&$-\ft5{48}(p+1)$\\
&&$\mathcal D(p+1,\ft12,\ft12;\ft23(p-2))(0,p-2)$&SLII&$-\ft5{48}(p+1)$\\
\hline
Gravitino I&$p=2$&$\mathcal D(\ft52,\ft12,0;\ft13)(1,0)$&conserved&$\ft{35}{192}$\\
&$p\ge2$&$\mathcal D(p+\ft12,\ft12,0;\ft23(p+\ft12))(0,p-1)$&chiral&$-\ft5{48}(p-1)$\\
&$p>2$& $\mathcal D(p+\ft12,\ft12,0;-\ft23(p-\ft52))(p-1,0)$&SLI&$-\ft1{96}(p+\ft12)$\\
&& $\mathcal D(p+\ft12,\ft12,0;\ft23(p-\ft32))(1,p-2)$&SLII&$\ft5{48}p$\\
\hline
Gravitino II&$p\ge3$&$\mathcal D(p+\ft32,\ft12,0;-\ft23(p-\ft32))(p-3,0)$&SLI&$-\ft1{96}(p+\ft32)$\\
\hline
Gravitino III&$p=2$&$\mathcal D(\ft52,0,\ft12;-\ft13)(0,1)$&conserved&$\ft{35}{192}$\\
&$p\ge2$ & $\mathcal D(p+\ft12,0,\ft12;-\ft23(p+\ft12))(p-1,0)$&anti-chiral&$-\ft5{48}(p-1)$\\
&$p>2$& $\mathcal D(p+\ft12,0,\ft12;\ft23(p-\ft52))(0,p-1)$&SLII&$-\ft1{96}(p+\ft12)$\\
&& $\mathcal D(p+\ft12,0,\ft12;-\ft23(p-\ft32))(p-2,1)$ & SLI&$\ft5{48}p$\\
\hline
Gravitino IV &$p\ge3$& $\mathcal D(p+\ft32,0,\ft12;\ft23(p-\ft32))(0,p-3)$&SLII&$-\ft1{96}(p+\ft32)$\\
\hline
Vector I &$p=2$&$\mathcal D(2,0,0;0)(1,1)$&conserved&$\ft1{32}$\\
&$p\ge2$&$\mathcal D(p,0,0;\ft23p)(0,p)$&chiral&$-\ft1{96}(p-\ft32)$\\
&&$\mathcal D(p,0,0;-\ft23p)(p,0)$&anti-chiral&$-\ft1{96}(p-\ft32)$\\
&$p>2$&$\mathcal D(p,0,0;-\ft23(p-2))(p-1,1)$&SLI&$\ft1{96}(p-\ft12)$\\
&&$\mathcal D(p,0,0;\ft23(p-2))(1,p-1)$&SLII&$\ft1{96}(p-\ft12)$\\
\hline
Vector II & --- & --- & --- &\\
\hline
Vector III &$p\ge2$& $\mathcal D(p+1,0,0;-\ft23(p+1))(p-2,0)$& anti-chiral&$-\ft1{96}(p-\ft12)$\\
&$p\ge3$& $\mathcal D(p+1,0,0;-\ft23(p-1))(p-3,1)$& SLI&$\ft1{96}(p+\ft12)$\\
\hline
Vector IV &$p\ge2$& $\mathcal D(p+1,0,0;\ft23(p+1))(0,p-2)$ & chiral&$-\ft1{96}(p-\ft12)$\\
&$p\ge3$& $\mathcal D(p+1,0,0;\ft23(p-1))(1,p-3)$ & SLII&$\ft1{96}(p+\ft12)$\\
\hline
\end{tabular}
\caption{\label{tbl:KKshort} Shortening structure of the $S^5$ KK
tower.  Note that Vector Multiplet II is never shortened.  The
contribution of a single shortened multiplet to $c-a$ is given in
the last column.  This factor must be multiplied by the dimension of
the SU(3) representation to obtain the total contribution to $c-a$.}
\end{table}

We are, of course, interested in $\mathbb Z_n$ orbifolds of $S^5$ generated
by the action of (\ref{eq:Znact}).  Since this element commutes with SU(2)
acting on the first two complex coordinates, it is natural to decompose the
original SU(4) $R$-symmetry according to
\begin{equation}
\mathrm{SU}(4)\supset\mathrm{SU}(3)\times\mathrm
U(1)_r\supset\mathrm{SU}(2)\times \mathrm U(1)_q\times\mathrm
U(1)_r. \label{eq:SU4toSU2U1U1}
\end{equation}
We define the U(1) normalizations by taking
\begin{equation}
\mathbf4\to\mathbf3_{1/3}+\mathbf1_{-1}\to\mathbf 2_{1,1/3}+\mathbf
1_{-2,1/3}+\mathbf 1_{0,-1}.
\end{equation}
Here the $R$-charge is conventionally normalized, while the U(1)$_q$
charge is normalized so that the states that survive the $\mathbb Z_{n}$
orbifolding are those that satisfy
\begin{equation}
q=0\mbox{ mod } n.
\label{eq:oproj}
\end{equation}
It is then simply a matter of group theory to project out the states
in the massive KK tower.

Before considering the orbifold, we rewrite the shortened $S^5$ spectrum in terms of
$\mathrm{SU}(2)\times\mathrm U(1)_q\times\mathrm U(1)_r$ quantum numbers.
This is obtained by appropriately branching the representations in Table~\ref{tbl:KKshort},
and the result is given in Table~\ref{tbl:SU2short}.  Of course this contains the same
information as Table~\ref{tbl:KKshort}.  However, it is now in a form that is applicable to
the $S^5/\mathbb Z_n$ orbifold models.

\begin{table}[t]
\centering
\begin{tabular}{|l|l|l|l|}
\hline
Multiplet&KK&Shortened representation&Shortening\\
&level&&type\\
\hline
Graviton&$p=2$&$\mathcal D(3,\ft12,\ft12;0)\mathbf 1_{0}$&conserved\\
&$p>2$&$\mathcal D(p+1,\ft12,\ft12;-\ft23(p-2))\sum_{k=0}^{p-2}(\mathbf{k+1})_{-2p+3k+4}$&SLI\\
&&$\mathcal D(p+1,\ft12,\ft12;\ft23(p-2))\sum_{k=0}^{p-2}(\mathbf{k+1})_{2p-3k-4}$&SLII\\
\hline
Gravitino I&$p=2$&$\mathcal D(\ft52,\ft12,0;\ft13)\mathbf1_{-2}+\mathbf2_{1}$&conserved\\
&$p\ge2$&$\mathcal D(p+\ft12,\ft12,0;\ft23(p+\ft12))\sum_{k=0}^{p-1}(\mathbf{k+1})_{2p-3k-2}$
&chiral\\
&$p>2$& $\mathcal D(p+\ft12,\ft12,0;-\ft23(p-\ft52))\sum_{k=0}^{p-1}(\mathbf{k+1})_{-2p+3k+2}$
&SLI\\
&& $\mathcal D(p+\ft12,\ft12,0;\ft23(p-\ft32))\sum_{k=1}^{p-1}(\mathbf{k+1})_{2p-3k}$&SLII \\
&&$\kern6em+\sum_{k=0}^{p-2}(\mathbf{k+1})_{2p-3k-6}$&\\
\hline
Gravitino II&$p\ge3$&$\mathcal D(p+\ft32,\ft12,0;-\ft23(p-\ft32))
\sum_{k=0}^{p-3}(\mathbf{k+1})_{-2p+3k+6}$&SLI\\
\hline
Gravitino III&$p=2$&$\mathcal D(\ft52,0,\ft12;-\ft13)\mathbf1_{2}+\mathbf2_{-1}$&conserved\\
&$p\ge2$ & $\mathcal
D(p+\ft12,0,\ft12;-\ft23(p+\ft12))\sum_{k=0}^{p-1}(\mathbf{k+1})_{-2p+3k+2}$
&anti-chiral \\
&$p>2$& $\mathcal D(p+\ft12,0,\ft12;\ft23(p-\ft52))\sum_{k=0}^{p-1}(\mathbf{k+1})_{2p-3k-2}$&SLII \\
&& $\mathcal D(p+\ft12,0,\ft12;-\ft23(p-\ft32))\sum_{k=1}^{p-1}(\mathbf{k+1})_{-2p+3k}$ & SLI \\
&&$\kern6em+\sum_{k=0}^{p-2}(\mathbf{k+1})_{-2p+3k+6}$&\\
\hline
Gravitino IV &$p\ge3$& $\mathcal D(p+\ft32,0,\ft12;\ft23(p-\ft32))
\sum_{k=0}^{p-3}(\mathbf{k+1})_{2p-3k-6}$&SLII\\
\hline
Vector I &$p=2$&$\mathcal D(2,0,0;0)\mathbf1_{0}+\mathbf2_{3}+\mathbf2_{-3}
+\mathbf3_{0}$&conserved\\
&$p\ge2$&$\mathcal D(p,0,0;\ft23p)\sum_{k=0}^p(\mathbf{k+1})_{2p-3k}$&chiral\\
&&$\mathcal D(p,0,0;-\ft23p)\sum_{k=0}^p(\mathbf{k+1})_{-2p+3k}$&anti-chiral\\
&$p>2$&$\mathcal D(p,0,0;-\ft23(p-2))\sum_{k=0}^{p-1}(\mathbf{k+1})_{-2p+3k+4}$&SLI\\
&&$\kern6em+\sum_{k=1}^p(\mathbf{k+1})_{-2p+3k-2}$&\\
&&$\mathcal D(p,0,0;\ft23(p-2))\sum_{k=1}^p(\mathbf{k+1})_{2p-3k+2}$&SLII\\
&&$\kern6em+\sum_{k=0}^{p-1}(\mathbf{k+1})_{2p-3k-4}$&\\
\hline
Vector II & --- & --- & --- \\
\hline
Vector III &$p\ge2$& $\mathcal D(p+1,0,0;-\ft23(p+1))
\sum_{k=0}^{p-2}(\mathbf{k+1})_{-2p+3k+4}$& anti-chiral \\
&$p\ge3$& $\mathcal
D(p+1,0,0;-\ft23(p-1))\sum_{k=0}^{p-3}(\mathbf{k+1})_{-2p+3k+8}$& SLI \\
&&$\kern6em+\sum_{k=1}^{p-2}(\mathbf{k+1})_{-2p+3k+2}$&\\
\hline
Vector IV &$p\ge2$& $\mathcal D(p+1,0,0;\ft23(p+1))
\sum_{k=0}^{p-2}(\mathbf{k+1})_{2p-3k-4}$ & chiral \\
&$p\ge3$& $\mathcal D(p+1,0,0;\ft23(p-1))\sum_{k=0}^{p-3}(\mathbf{k+1})_{2p-3k-8}$& SLII \\
&&$\kern6em+\sum_{k=1}^{p-2}(\mathbf{k+1})_{2p-3k-2}$&\\
\hline
\end{tabular}
\caption{\label{tbl:SU2short} The shortened multiplets of the $S^5$ KK tower decomposed in
terms of $\mathrm{SU}(2)\times\mathrm U(1)_q\times\mathrm U(1)_r$ quantum numbers.
Note that the SU(2) representations are given in terms of their dimensions.}
\end{table}

\subsection{Subleading Weyl anomaly computation}

We now turn to the computation of $c-a$ for the orbifolds
$S^5/\mathbb Z_n$. Basically, our goal is to sum the individual
contributions given in the last column of Table~\ref{tbl:KKshort}
over the shortened representations of Table~\ref{tbl:SU2short} that
survive the orbifolding.  It is more convenient to rewrite the sums
over KK level $p$ and SU(2) representation $k$ in
Table~\ref{tbl:SU2short} in terms of sums over the U(1)$_q$ charge
$q$ and KK level $p$.  In this case, we can then restrict the sum
over $q$ to those satisfying the projection condition
(\ref{eq:oproj}), namely $q=0\mbox{ mod }n$.

One simplifying step is to note that the contribution to $c-a$ from the conserved
multiplets at KK level $p=2$ in fact matches the sum of the corresponding
contributions to $c-a$ from the SLI and SLII multiplets, if their contributions were to
be extrapolated from $p>2$ to $p=2$.  For example, if we took the graviton SLI
and SLII contributions from Table~\ref{tbl:KKshort} and set $p=2$, we would find
\begin{equation}
-\fft5{48}(p+1)-\fft5{48}(p+1)\Big|_{p=2}=-\fft58,
\end{equation}
which agrees with the value for the conserved graviton multiplet.  It is easy to see
that this holds in general for all of the conserved multiplets.

Continuing with the graviton multiplet, since the SLI and SLII multiplets are conjugates
of each other, it is sufficient to consider only one of them, and double the result.  We
are thus led to the contribution
\begin{equation}
c-a\Big|_{\rm graviton}=2\times\sum_{p=2}^\infty\sum_{k=0}^{p-2}z^p\left(-\fft5{48}\right)(p+1)(k+1),
\label{eq:exsum}
\end{equation}
where $z^p$ is used to regulate the sum over KK modes, and where the $k+1$ factor is
the dimension of the SU(2) representation.  For the $\mathbb Z_n$ orbifold, the sum
in (\ref{eq:exsum}) should be restricted to $q=0\mbox{ mod }n$, where $q=-2p+3k+4$.

In order to make the $q$ charge explicit, we write
\begin{equation}
c-a\Big|_{\rm graviton}=\sum_{p=2}^\infty\sum_{k=0}^{p-2}f(-2p+3k+4,p,k),
\end{equation}
where $f(q,p,k)$ is the summand in (\ref{eq:exsum}).  It is a simple exercise to
convert this into a set of sums over $q$
\begin{eqnarray}
c-a\Big|_{\rm graviton}&=&\sum_{j=0}^\infty\sum_{l=0}^\infty f(j,j+3l+2,j+2l)
+\sum_{j=1}^\infty\sum_{l=0}^\infty f(-2j,j+3l+2,2l)\nn\\
&&+\sum_{j=0}^\infty\sum_{l=0}^\infty f(-2j-1,j+3l+4,2l+1).
\label{eq:threesums}
\end{eqnarray}
In particular, the first sum in (\ref{eq:threesums}) is over non-negative $q$, the
second sum is over negative even $q$ and the final sum is over negative odd $q$.
Given this decomposition, it is now straightforward to restrict the $q$ charges for
the $\mathbb Z_n$ orbifold.

Note that for even $n$, the negative odd $q$ sum in (\ref{eq:threesums}) drops out,
while for odd $n$ all three sums will contribute.  Thus we consider even and odd cases
separately.  For even $n$, we have
\begin{equation}
c-a\Big|_{\rm graviton}^{{\rm even~} \mathbb Z_n}=\sum_{j=0}^\infty\sum_{l=0}^\infty f(nj,nj+3l+2,nj+2l)
+\sum_{j=1}^\infty\sum_{l=0}^\infty f(-nj,nj/2+3l+2,2l),
\end{equation}
and for odd $n$ we have
\begin{eqnarray}
c-a\Big|_{\rm graviton}^{{\rm odd~}\mathbb Z_n}&=&
\sum_{j=0}^\infty\sum_{l=0}^\infty f(nj,nj+3l+2,nj+2l)
+\sum_{j=1}^\infty\sum_{l=0}^\infty f(-2nj,nj+3l+2,2l)\nn\\
&&+\sum_{j=0}^\infty\sum_{l=0}^\infty f(-n(2j+1),(n(2j+1)-1)/2+3l+4,2l+1).
\end{eqnarray}
In both cases, the function $f(q,p,k)$ for the graviton is given in (\ref{eq:exsum}):
\begin{equation}
f(q,p,k)=-\fft5{24}z^p(p+1)(k+1).
\end{equation}
The sums can be evaluated, and the result for the graviton contribution is
\begin{equation}
c-a\Big|_{\rm graviton}=\begin{cases}\displaystyle
-\fft5{8n(z-1)^4}-\fft5{4n(z-1)^3}-\fft{65}{96n(z-1)^2}+\fft{n^4+20n^2+84}{4608n}+\cdots,&
n\mbox{ even};\kern-3pt\\
\displaystyle
-\fft5{8n(z-1)^4}-\fft5{4n(z-1)^3}-\fft5{8n(z-1)^2}+\fft{n^4+30n^2-31}{4608n}+\cdots,&
n\mbox{ odd}.
\end{cases}
\end{equation}
Recall that $z$ is used to regulate the sum over the infinite KK tower; following
\cite{Mansfield:2002pa}, we expect to ignore the pole terms and keep only the
finite contribution to $c-a$.

To obtain the full result, we sum over all shortened multiplets in
Table~\ref{tbl:SU2short}.  Since the procedure for the other
multiplets parallels that of the graviton multiplet, we omit the
details here.  However, there is one small detail for some of the
other multiplets, which is that the restriction on KK level leads to
a few exceptions in the sums for $n=1$ (ie the round $S^5$) and
$n=2$ (ie $S^5/\mathbb Z_2$).  These exceptions are perhaps not
surprising, as these cases have additional supersymmetry compared
with the generic orbifolds.

\subsubsection{Odd orbifolds}

For odd $n$, the $\mathbb Z_n$ element (\ref{eq:Znact}) acts freely on $S^5$.
Hence there is no need to consider any twisted sectors, and the sum over the shortened
KK spectrum gives the entire contribution to $c-a$.  Curiously, the pole terms
vanish identically when summing over all multiplets, and we are left with the simple
result
\begin{equation}
c-a\Big|_{S^5/\mathbb Z_n}=\begin{cases}0,&n=1;\\
\displaystyle\fft{n}{16}+\cdots,&n\ge3\mbox{ odd},
\end{cases}
\end{equation}
where the ellipses denote terms vanishing in the limit $z\to1$.
This matches the field theory result (\ref{eq:ftresult}).

\subsubsection{Even orbifolds}

For even $n$, there is the added complication that the $\mathbb Z_n$ action
admits a $\mathbb Z_2$ subgroup generated by
\begin{equation}
\Omega^{n/2}=\begin{pmatrix}-1\\&-1\\&&\hphantom{-}1\end{pmatrix}.
\label{eq:Z2act}
\end{equation}
This element leaves a fixed plane in $\mathbb C^3$, which gives rise
to a fixed circle on $S^5$.  Thus, to understand the even orbifolds,
we will have to consider the effect of the twisted sector in
addition to the KK tower discussed above.

\begin{figure}[t]
\centering
    \includegraphics[scale=.2]{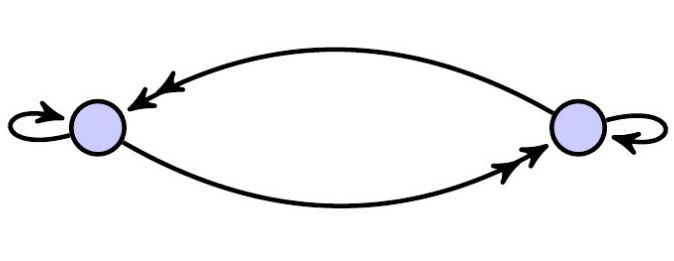}
    \hspace{2cm}
    \includegraphics[scale=.25]{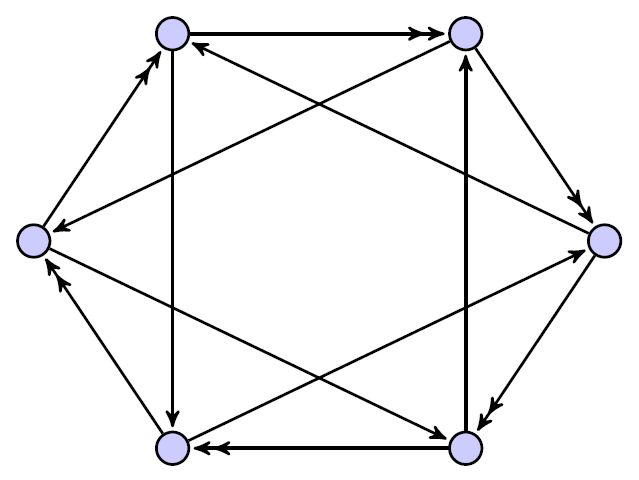}
\caption{Two of the orbifold quivers. The $\mathbb{Z}_{2}$ orbifold is special because
of the chiral multiplets in the adjoints and is shown on the left. The $\mathbb{Z}_{6}$
orbifold follows the generic pattern and is shown on the right. \label{fig:orbiquivers}}
\end{figure}

Before discussing the twisted sector, we present the result from the sum
over the shortened KK spectrum in the untwisted sector
\begin{equation}
c-a\Big|_{\rm untwisted}=\begin{cases}\displaystyle
-\fft1{8(z-1)^2}-\fft1{8(z-1)}+\fft1{16}+\cdots,&n=2;\\
\displaystyle
-\fft1{4n(z-1)^2}-\fft1{4n(z-1)}+\fft{5n^2-4}{96n}+\cdots,&n\ge4\mbox{ even}.
\end{cases}
\label{eq:evenuntwisted}
\end{equation}
Unlike in the odd case, here the pole terms do not disappear.  Note, however, that
the leading fourth and third order poles cancel when summed over the complete
set of multiplets.  The $n=2$ case is an exception since the dual quiver gauge theory
has $\mathcal N=2$ supersymmetry and chiral matter in the adjoint, as indicated in
Figure~\ref{fig:orbiquivers}.

The twisted modes of the even orbifolds are known to arise from the
KK reduction of the six-dimensional $(2,0)$ tensor theory on AdS$_5\times
S^1$ \cite{Gukov:1998kk,Gadde:2009dj}.  Since these states originate
from the $\mathbb Z_2$ action generated by (\ref{eq:Z2act}), they
preserve $\mathcal N=4$ supersymmetry in five dimensions.  However,
they may be further decomposed into $\mathcal N=2$ multiplets.  The
result is presented in Table~\ref{tbl:twisted}.  (See Appendix~\ref{app:twisted}
for additional details.)  For even $n\ge4$,
an additional projection $q=0\mbox{ mod }n$ must be imposed. In
this case, the zero mode ($p=0$) must be treated separately from the
KK tower on $S^1$.  We find
\begin{equation}
c-a\Big|_{p=0~\rm
twisted}=\begin{cases}1/48,&n=2;\\1/32,&n\ge4\mbox{
even},\end{cases} \label{eq:evenzero}
\end{equation}
and
\begin{equation}
c-a\Big|_{p\ge1~\rm twisted}=\begin{cases}\displaystyle
-\fft1{4(z-1)^2}-\fft1{4(z-1)}+0+\cdots,&n=2;\\
\displaystyle
-\fft1{2n(z-1)^2}-\fft1{2n(z-1)}+\fft{n^2-3n+4}{96n},&n\ge4\mbox{ even}.
\end{cases}
\label{eq:eventwisted}
\end{equation}
Again, a $z^{p}$ regulator is used, with $p$ the KK level on
$S^{1}$. Here the double pole is leading, so there is no partial
pole cancellation as there was in the untwisted sector. Note that
one could introduce a different fugacity for the twisted states,
say multiplying each term by $y^p$ instead of $z^p$, but that
would not change the finite part of the final result in
(\ref{eq:evenfinal}) after one also expands around $y=1$.

\begin{table}[t]
\centering
\begin{tabular}{|l|l|l|}
\hline
KK level&Representation&Shortening type\\
\hline
$p=0$&$\mathcal D(2,0,0;0)\mathbf1_0$&conserved\\
&$\mathcal D(2,0,0;\ft43)\mathbf1_{-2}$&chiral\\
&$\mathcal D(2,0,0;-\ft43)\mathbf1_2$&anti-chiral\\
$p\ge1$&$\mathcal D(p+1,0,0;\ft23(p+1))\mathbf1_{2p+2}$&chiral\\
&$\mathcal D(p+\ft32,\ft12,0;\ft23(p+\ft32))\mathbf1_{2p}$&\\
&$\mathcal D(p+2,0,0;\ft23(p+2))\mathbf1_{2p-2}$&\\
&$\mathcal D(p+1,0,0;-\ft23(p+1))\mathbf1_{-2p-2}$&anti-chiral\\
&$\mathcal D(p+\ft32,0,\ft12;-\ft23(p+\ft32))\mathbf1_{-2p}$&\\
&$\mathcal D(p+2,0,0;-\ft23(p+2))\mathbf1_{-2p+2}$&\\
\hline
\end{tabular}
\caption{\label{tbl:twisted}The twisted sector states for the orbifold $S^5/\mathbb Z_2$
written in an $\mathcal N=2$ language.  We use the same $\mathrm{SU}(2)\times\mathrm U(1)_q
\times\mathrm U(1)_r$ decomposition as in (\ref{eq:SU4toSU2U1U1}).}
\end{table}

Adding together (\ref{eq:evenuntwisted}), (\ref{eq:evenzero}) and (\ref{eq:eventwisted}),
we find
\begin{equation}
c-a\Big|_{S^5/\mathbb Z_n}=\begin{cases}\displaystyle
-\fft3{8(z-1)^2}-\fft3{8(z-1)}+\fft1{12}+\cdots,&n=2;\\
\displaystyle
-\fft3{4n(z-1)^2}-\fft3{4n(z-1)}+\fft{n}{16}+\cdots,&n\ge4\mbox{ even}.
\end{cases}
\label{eq:evenfinal}
\end{equation}
Although the second and first order poles survive in this result, if we follow the
regulation procedure of \cite{Mansfield:2002pa} and drop the poles, we see that
the finite part agrees with the field theory result (\ref{eq:ftresult}).

Thus we have successfully reproduced the field theory result for $c-a$, (\ref{eq:ftresult}),
for all $\mathbb Z_n$ orbifolds of $S^5$.  For odd $n$, the regulated sum over
the KK tower is finite, and directly gives $c-a=n/16$.  For even $n$, the regulated sum
diverges with first and second order poles.  However, the finite term correctly gives $c-a=n/16$
(or $c-a=1/12$ for $n=2$).  This distinction between even and odd orbifolds is presumably
related to the presence of a twisted sector in the former case.  Furthermore, the holographic
contribution to $c-a$ from massive string loops vanishes in this case \cite{Liu:2010gz}, so the result
from the KK tower is complete.

\section{Orbifolds of $T^{1,1}$}
\label{sec:T11}

Having successfully matched the gravity and field theory results for $c-a$ for
the supersymmetric orbifolds $S^5/\mathbb Z_n$, we would like to extend this
comparison to more quiver gauge theories and their gravitational duals.  Because
the holographic computation requires knowledge of the shortened KK spectrum,
we restrict our consideration to $T^{1,1}$, where the spectrum is known
\cite{Ceresole:1999zs,Ceresole:1999ht}.

\subsection{The spectrum and shortenings}

As demonstrated in \cite{Eager:2012hx}, the generic KK spectrum for
compactification of IIB supergravity on a Sasaki-Einstein manifold
consists of nine generic KK multiplets (originally identified for
$T^{1,1}$ in \cite{Ceresole:1999zs,Ceresole:1999ht}), along with
possibly additional `special' KK multiplets and Betti multiplets. An
example of the special and Betti multiplets can be seen in the case
of $S^5/\mathbb Z_2$, were the twisted sector states shown in
Table~\ref{tbl:twisted} can be organized into three chiral and three
anti-chiral towers, corresponding to special multiplets, along with
the three $q=0$ representations $\mathcal D(2,0,0;0)$, $\mathcal
D(3,0,0,2)$ and $\mathcal D(3,0,0,-2)$, corresponding to Betti
multiplets\footnote{Recall that the topology of
$S^{5}/\mathbb{Z}_{2}$ with the fixed circle blown up is the same as
$T^{1,1}$ and therefore it admits the same type of Betti multiplets
as the latter.}. These special and Betti multiplets do not exist for
the round $S^5$ nor for its odd orbifolds.

For a given Sasaki-Einstein manifold, the KK spectrum (excluding special and Betti multiplets)
can be obtained in terms of the eigenvalues of the scalar Laplacian
\cite{Ceresole:1999zs,Ceresole:1999ht,Eager:2012hx}.  For $T^{1,1}$, define
the eigenvalues of the scalar Laplacian on $T^{1,1}$ as
\begin{equation}
\square Y = -H_0 Y,\qquad H_0=6[j(j+1)+\ell(\ell+1)-r^2/8],
\end{equation}
where $(j,\ell,r)$ specify the
$\mathrm{SU}(2)\times\mathrm{SU}(2)\times\mathrm U(1)$ quantum
numbers.  Here the $r$-charge is integer quantized, and is bounded
by
\begin{equation}
|r|\le2\min(j,\ell).
\end{equation}
Now let
\begin{equation}
H_0=e_0(e_0+4),
\end{equation}
or
\begin{equation}
e_0=\sqrt{H_0+4}-2\ge0.
\end{equation}
The Kaluza-Klein supermultiplet spectrum on $T^{1,1}$ is then given
in Table~\ref{tbl:T11N=2spectrum}. Note that $e_0=0$ corresponds to
the zero mode on $T^{1,1}$, with $j=\ell=r=0$.  There are four sets
of supermultiplets where this is allowed; these are the ones that
may be retained in the massive consistent truncation on
Sasaki-Einstein
\cite{Cassani:2010uw,Liu:2010sa,Gauntlett:2010vu,Skenderis:2010vz},
and they are shown in Table~\ref{tbl:T11KKconsistent}.

\begin{table}[tp]
\centering
\begin{tabular}{|l|l|l|}
\hline
Supermultiplet&Representation&$e_0$ condition\\
\hline
Graviton&$\mathcal D(e_0+3,\ft12,\ft12;r)$&$e_0\ge0$\\
Gravitino I and III&$\mathcal D(e_0+\ft32,\ft12,0;r+1)+\mathcal
D(e_0+\ft32,0,\ft12;r-1)$&
$e_0>0$\\
Gravitino II and IV&$\mathcal D(e_0+\ft92,\ft12,0;r-1)+\mathcal
D(e_0+\ft92,0,\ft12;r+1)$&
$e_0\ge0$\\
Vector I&$\mathcal D(e_0,0,0;r)$&$e_0>0$\\
Vector II&$\mathcal D(e_0+6,0,0;r)$&$e_0\ge0$\\
Vector III and IV&$\mathcal D(e_0+3,0,0;r-2)+\mathcal D(e_0+3,0,0;r+2)$&$e_0\ge0$\\
\hline
Betti vector&$\mathcal D(2,0,0;0)$&\\
Betti hyper&$\mathcal D(3,0,0;2)+\mathcal D(3,0,0;-2)$&\\
\hline
\end{tabular}
\caption{\label{tbl:T11N=2spectrum} The $\mathcal N=2$ spectrum of
IIB supergravity on $T^{1,1}$.  All representations transform as
$(j,\ell)$ under $\mathrm{SU}(2)\times\mathrm{SU}(2)$.}
\end{table}

\begin{table}[tp]
\centering
\begin{tabular}{|l|l|l|}
\hline
Supermultiplet&Representation&Name given in \cite{Liu:2010sa}\\
\hline
Graviton&$\mathcal D(3,\ft12,\ft12;0)$&supergraviton\\
Gravitino II and IV&$\mathcal D(\ft92,\ft12,0;-1)+\mathcal D(\ft92,0,\ft12,1)$
&LH+RH massive gravitino\\
Vector II&$\mathcal D(6,0,0;0)$&massive vector\\
Vector III and IV&$\mathcal D(3,0,0;-2)+\mathcal D(3,0,0;2)$&LH+RH chiral\\
\hline
\end{tabular}
\caption{\label{tbl:T11KKconsistent} The $e_0=0$ multiplets.  These
are the multiplets that survive the consistent Sasaki-Einstein
truncation.}
\end{table}

It is straightforward to work out the multiplet shortening conditions for the
$T^{1,1}$ spectrum, and the result is shown in Table~\ref{tbl:T11KKshort}.
We have also included the Betti multiplets in this table, as they are
part of the shortened spectrum.  There are no special multiplets for $T^{1,1}$.

Although the $T^{1,1}$ harmonics do not
have an obvious single `KK level' arrangement (since they involve harmonics
on $S^2\times S^2\times S^1$ instead of a single $S^5$), the shortened
multiplets follow the same pattern as those of $S^5$. Thus in
Table~\ref{tbl:T11KKshort} we have assigned KK levels based on what
they would have been for $S^5$ (or its orbifolds).  Note, however,
that $j$ can take on both integer and half-integer values.  The
lowest KK level is $p=3/2$, and it consists of only Vector I (chiral and
anti-chiral).  In fact, this lowest KK level is unusual, in that the shortened
representation $\mathcal D(\fft32,0,0;1)(\fft12,\fft12) +
\mathcal D(\fft32,0,0;-1)(\fft12,\fft12)$ contains a complex scalar with $E_0=3/2$.
This is in the range where both modes are normalizable.  Thus the
scalar needs to be quantized with Neumann (as opposed to the usual
Dirichlet) boundary conditions in order to select out the $E_0=3/2$ mode.
We will have more to say more about this below.

\begin{table}[tp]
\centering
\begin{tabular}{|l|l|l|l|}
\hline
Multiplet&KK level&Shortened representation&Shortening type\\
\hline
Graviton&$p=2$&$\mathcal D(3,\ft12,\ft12;0)(0,0)$&conserved\\
&$p=3j+2$&$\mathcal D(3j+3,\ft12,\ft12;-2j)(j,j)$&SLI ($j>0$)\\
&&$\mathcal D(3j+3,\ft12,\ft12;2j)(j,j)$&SLII ($j>0$)\\
\hline
Gravitino I&$p=3j+1$&$\mathcal D(3j+\ft32,\ft12,0;2j+1)(j,j)$&chiral ($j>0$)\\
&& $\mathcal D(3j+\ft32,\ft12,0;-2j+1)(j,j)$&SLI ($j>0$)\\
&$p=3j+3$& $\mathcal D(3j+\ft72,\ft12,0;2j+1)(j+1,j)\oplus(j,j+1)$&SLII \\
\hline
Gravitino II&$p=3j+3$&$\mathcal D(3j+\ft92,\ft12,0;-2j-1)(j,j)$&SLI\\
\hline
Gravitino III&$p=3j+1$ & $\mathcal D(3j+\ft32,0,\ft12;-2j-1)(j,j)$&anti-chiral ($j>0$)\\
&& $\mathcal D(3j+\ft32,0,\ft12;2j-1)(j,j)$&SLII ($j>0$)\\
&$p=3j+3$& $\mathcal D(3j+\ft72,0,\ft12;-2j-1)(j+1,j)\oplus(j,j+1)$ & SLI\\
\hline
Gravitino IV &$p=3j+3$& $\mathcal D(3j+\ft92,0,\ft12;2j+1)(j,j)$&SLII\\
\hline
Vector I &$p=2$&$\mathcal D(2,0,0;0)(1,0)\oplus(0,1)$&conserved\\
&$p=3j$&$\mathcal D(3j,0,0;2j)(j,j)$&chiral ($j>0$)\\
&&$\mathcal D(3j,0,0;-2j)(j,j)$&anti-chiral ($j>0$)\\
&$p=3j+2$&$\mathcal D(3j+2,0,0;-2j)(j+1,j)\oplus(j,j+1)$&SLI ($j>0$)\\
&&$\mathcal D(3j+2,0,0;2j)(j+1,j)\oplus(j,j+1)$&SLII ($j>0$)\\
\hline
Vector II & --- & --- & --- \\
\hline
Vector III &$p=3j+2$& $\mathcal D(3j+3,0,0;-2j-2)(j,j)$& anti-chiral \\
&$p=3j+4$& $\mathcal D(3j+5,0,0;-2j-2)(j+1,j)\oplus(j,j+1)$& SLI \\
\hline
Vector IV &$p=3j+2$& $\mathcal D(3j+3,0,0;2j+2)(j,j)$ & chiral \\
&$p=3j+4$& $\mathcal D(3j+5,0,0;2j+2)(j+1,j)\oplus(j,j+1)$ & SLII \\
\hline
Betti vector&$-$&$\mathcal D(2,0,0;0)(0,0)$&conserved\\
\hline
Betti hyper&$-$&$\mathcal D(3,0,0;2)(0,0)$&chiral\\
&&$\mathcal D(3,0,0;-2)(0,0)$&anti-chiral\\
\hline
\end{tabular}
\caption{Shortening structure of the
$T^{1,1}$ KK tower. The supermultiplets are given in the
conventional notation $\mathcal D(E_0,s_1,s_2;r)$ with the
$\mathrm{SU}(2)\times\mathrm{SU}(2)$ representation $(j,\ell)$
appended.  Here $j=0,\ft12,1,\ft32,\ldots$, unless otherwise indicated.
Note that Vector Multiplet II is never shortened.  The `KK
level' is suggested by analogy with the $S^5$ spectrum.
\label{tbl:T11KKshort}}
\end{table}

\subsection{Subleading Weyl anomaly computation}

The holographic computation of $c-a$ proceeds along the same
lines as that for $S^5$.  We essentially take the contributions to
$c-a$ from Table~\ref{tbl:KKshort} and sum over the shortened
$T^{1,1}$ spectrum of Table~\ref{tbl:T11KKshort}.  As in the $S^5$
case, we can simplify the sum over the spectrum by ignoring the
conserved Graviton and Vector I multiplets and instead extend
the sums for the corresponding SLI and SLII towers to include
$j=0$.

As an example, we present the computation of $c-a$ for the
graviton tower.  Using the same regularization procedure of
multiplying by $z^p$ (where $p$ is the assigned KK level), we have
\begin{eqnarray}
c-a\Big|_{\rm
graviton}&=&2\times\sum z^{p}\left(-\frac{5}{48}\right)(p+1)(2j+1)(2l+1)\nn \\
&=&2\times\sum_{j}z^{3j+2}\left(-\frac{5}{48}\right)(3j+3)(2j+1)^2,
\label{eq:c-aT11graviton}
\end{eqnarray}
where $j=0,\frac{1}{2},1,\ldots$, and the overall factor of two
takes care of the conjugate multiplets.  In the first line, the factor $(2j+1)(2l+1)$
corresponds to the dimension of the $\mathrm{SU}(2)\times\mathrm{SU}(2)$
representation, and in the second line we have substituted in the relation
between $p$, $j$ and $l$ as shown in Table~\ref{tbl:T11KKshort}. The sum
can be easily evaluated, and the result for the graviton contribution is
\begin{equation}
c-a\Big|_{\rm graviton}=-\frac{10}{27(z-1)^{4}}-\frac{20}{27(z-1)^{3}}
-\frac{125}{324(z-1)^{2}}+\frac{385}{31104}+\cdots.
\end{equation}

The contributions from the other towers can be worked out in a similar manner.
In addition, the contribution from the Betti vector ($1/32$) cancels against that
from the Betti hyper ($-1/32$).  There is one subtlety, however, and that is
related to the quantization of the $E_0=3/2$ scalar in the $p=3/2$ KK level,
as mentioned above.  The alternate boundary conditions used to quantize
this scalar may modify its contribution to $c-a$ \cite{Nolland:2003kc}.  Thus
we add a term $\delta_{\rm alt.~quant.}$ that accounts for this contribution.
This, however, is at most a finite shift, and will not affect the convergence
of the sum over the KK tower.
Putting everything together, one arrives at
\begin{equation}
c-a\Big|_{T^{1,1}}=-\fft2{9(z-1)^2}-\fft2{9(z-1)}+\fft1{8}+\delta_{\rm alt.~quant.}+\cdots.
\label{eq:T11ans}
\end{equation}
While the fourth and third order poles cancel at $z=1$, the second
and first order poles do not. Hence the sum over the KK tower is
divergent.  Following the prescription of \cite{Mansfield:2002pa}, we
drop the pole terms, so we are left with the finite result
$c-a=1/8+\delta_{\rm alt.~quant.}$ for $T^{1,1}$.

The shift $\delta_{\rm alt.~quant.}$ due to imposing Neumann boundary
conditions for the $E_0=3/2$ scalar is not well understood.
Ref.~\cite{Nolland:2003kc} reports an answer for this shift corresponding
to $\delta=-1/180$ for each real scalar with $E_0<2$,
but finds disagreement with the established results
\cite{Gubser:2002zh,Gubser:2002vv,Hartman:2006dy} on the shift in
the $a$ central charge due to the alternative boundary condition.
We are not able to resolve this contradiction.  However, it is interesting to
note that the conifold gauge theory has $c-a=1/8$, suggesting that
$\delta_{\rm alt.~quant.}$ should in fact vanish.

\subsubsection{The $T^{1,1}/\mathbb{Z}_{2}$ orbifold}
\label{sec:T11modZ2}

\begin{figure}[t]
\centering
    \includegraphics[scale=.2]{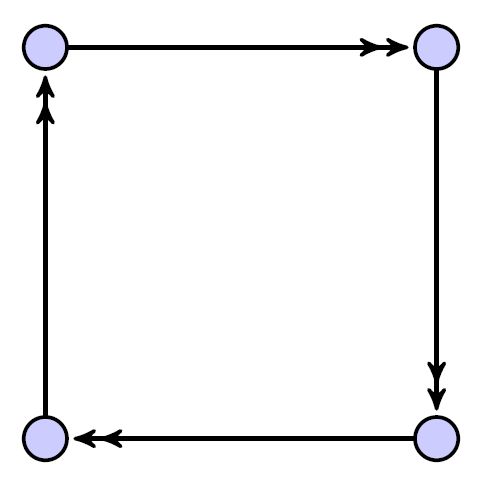}
    \hspace{2cm}
    \includegraphics[scale=.3]{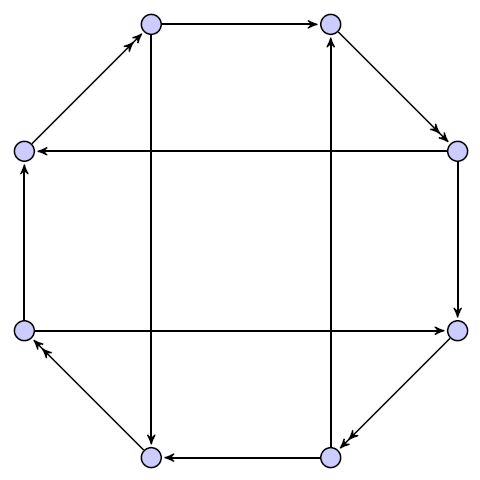}
\caption{The quivers corresponding to $T^{1,1}/\mathbb{Z}_{2}$ (on the left)
and $T^{1,1}/\mathbb{Z}_{4}$ (on the right). The latter is an example of typical
quivers corresponding to orbifolds of $T^{1,1}$. \label{T11modZ2quiver}}
\end{figure}

One way to avoid the issue of working with alternate boundary conditions is
to consider orbifolds of $T^{1,1}$ where the $p=3/2$ KK level is projected
out.  We first consider the orbifold $T^{1,1}/\mathbb Z_2$ defined by taking
the period along the U(1) fiber to be $2\pi$ instead of the normal $4\pi$.
This orbifold maintains the $\mathrm{SU}(2)\times\mathrm{SU}(2)\times\mathrm U(1)$
isometry of $T^{1,1}$, but projects to integer SU(2) charges only.
This corresponds to taking integer $j$ in Table~\ref{tbl:T11KKshort},
so the KK level $p$ is now an integer; in particular this removes the
$p=3/2$ multiplets from the spectrum.
The dual quiver is shown in Figure~\ref{T11modZ2quiver}.

Computing the holographic $c-a$ is quite similar to the case of
$T^{1,1}$, but the sums are now over integer $j$. We find
\begin{equation}
c-a\Big|_{T^{1,1}/\mathbb{Z}_{2}}=-\fft1{9(z-1)^2}-\fft1{9(z-1)}+\fft1{4}+\cdots.
\end{equation}
Keeping the finite part gives $c-a=1/4$, in perfect agreement with the
field theory result corresponding to the four-node dual quiver.

\subsubsection{The $T^{1,1}/\mathbb Z_n$ orbifolds}

We now consider the $Y^{n,0}=T^{1,1}/\mathbb Z_n$ orbifolds obtained by taking
a $\mathbb Z_n$ quotient of the conifold.  In particular, we take the conifold to
be defined by
\begin{equation}
xy-zw=0.
\end{equation}
Then the $\mathbb{Z}_{n}$ action is defined by \cite{Franco:2005fd}
\begin{equation}
x\rightarrow e^{2\pi i/n}x,\quad y\rightarrow e^{-2\pi i/n}y,
\quad z\rightarrow e^{2\pi i/n}z,\quad w\rightarrow e^{-2\pi i/n}w.
\end{equation}
This corresponds to a $\mathbb Z_n$ quotient of the $\mathrm{SU}(2)_j$
subgroup of the isometry group $\mathrm{SU}(2)_j\times\mathrm{SU}(2)_l
\times\mathrm U(1)_r$ of $T^{1,1}$.  Note that these orbifolds are all
fixed-point-free, and that the $n=2$ case corresponds to taking integer $j$,
and hence reduces to the $\mathbb Z_2$ orbifold considered above.
For $n>2$, the isometry group of the orbifold is reduced to $\mathrm{SU}(2)_l
\times\mathrm U(1)_j\times\mathrm U(1)_r$.

To find the $\mathbb{Z}_{n}$-singlet states, one simply decomposes
$\mathrm{SU}(2)_j\supset \mathrm U(1)_j$, where $\mathrm U(1)_j$ is
just the third component of isospin, and then keeps $j_z=0\mbox{ mod
}n/2$. For example, the conserved graviton multiplet with
$(j,l)=(0,0)$ along with the conserved vector multiplet with
$(j,l)=(0,1)$ survives the orbifolding. However, the conserved
vector with $(j,l)=(1,0)$ will be branched to $(0)_{-1}+(0)_0+(0)_1$
(where the $l$ quantum number is shown inside the parentheses and
the $\mathrm U(1)_j$ charge is subscripted). Only the $(0)_0$ state
will survive the $\mathbb Z_n$ projection for $n>2$.

As above, we highlight the computation of $c-a$ for the graviton tower.  We have
\begin{equation}
c-a\Big|_{\rm
graviton}=2\times\sum_{j}z^{p}\left(-\frac{5}{48}\right)(p+1)
\gamma^{(n)}_{j}(2l+1).
\end{equation}
This expression is identical to the first line of
(\ref{eq:c-aT11graviton}), except that the dimension of the complete
$\mathrm{SU}(2)_j$ representation, $2j+1$, is replaced by
$\gamma^{(n)}_{j}$, which counts the number of states surviving the
orbifolding by $\mathbb Z_n$.  For example, take $n=3$ and consider
the $\mathrm{SU}(2)_j$ representation given by $j=4$.  The $j_z$
charges are then all integers from $-4$ to $4$, and only three of
the states, with $j_z=-3,0,3$ survive the projection.  Hence we find
$\gamma^{(3)}_{4}=3$.  For the general case, we may write
$j=(n\alpha+\beta)/2$, with $\alpha$, $\beta$ nonnegative integers
where $\beta<n$. When $n$ is even, it turns out that
$\gamma^{(n)}_{j}=2\alpha+1$. When $n$ is odd,
$\gamma^{(n)}_{j}=\alpha$ for odd $\beta$, and
$\gamma^{(n)}_{j}=\alpha+1$ for even $\beta$.

We skip the rest of the details and report the final answer
\begin{equation}
c-a\Big|_{T^{1,1}/\mathbb{Z}_{n}}=-\fft2{9n(z-1)^2}-\fft2{9n(z-1)}+\fft
n{8}+\cdots.
\end{equation}
Interestingly, although the computation bifurcates depending on even or odd $n$,
this final result takes the same form in both cases.  Setting $n=1$ reproduces
the $T^{1,1}$ answer (\ref{eq:T11ans}), but without $\delta_{\rm alt.~quant.}$.
Keeping only the finite part, we find $c-a=n/8$, again in agreement with the
field theory result for the $2n$-node quiver corresponding to $Y^{n,0}$.

It appears that we have been successful in reproducing the quiver field theory
result $c-a=n/8$ for the entire family of $\mathbb Z_n$ orbifolds of $T^{1,1}$.
However, this does raise a puzzle in that it was argued in \cite{Liu:2010gz} that $c-a$
for $T^{1,1}$ would receive an additional contribution of $1/24$ from massive
string loop corrections, and it can be shown that this corresponds to a
contribution of $n/24$ for $T^{1,1}/\mathbb Z_n$.  For $T^{1,1}$ itself, this
would suggest that $\delta_{\rm alt.~quant.}$ in (\ref{eq:T11ans}) should
take the value $-1/24$, so as to cancel the massive string loop correction.  However,
there is no added room for removing the $n/24$ contribution for the orbifolds
with $n>1$.  This suggests that the conjecture in \cite{Liu:2010gz} that one simply adds
the massive string loop to the supergravity KK loop contributions in order
to obtain $c-a$ needs refinement.

\section{Discussion}
\label{sec:dis}

Our main result is the exact matching of the holographic $c-a$ with
the gauge theory result for the families of theories dual to IIB
string theory on $S^5/\mathbb Z_n$ and $T^{1,1}/\mathbb Z_n$.  This
exact matching is achieved by considering in the bulk effective
supergravity theory the one-loop contribution%
\footnote{We need $\alpha'$ and $g_{s}$ to be small enough (or
$\lambda$ and $N$ large enough) that the ten dimensional
supergravity gives a good approximation to the bulk effective
action, but we must not take strict limits, as it would hide the
subleading effects due to respectively massive and massless loops.
The fact that $c-a$ of the quivers are independent of both $\lambda$
and $N$ then seems to guarantee the one-loop exactness of the
gravitational results.}
on AdS$_5$ with all possible states in the shortened KK spectrum
running in the loop. (The long multiplets have a vanishing
contribution, and hence can be discarded from the computation.)
Since the KK tower is unbounded, the sum (\ref{eq:c-aexpr}) over the
tower does not converge, and needs to be regulated.  We have
followed the regularization method of \cite{Mansfield:2003gs}, which
is to multiply the contribution at each KK level by $z^p$, where $p$
is the level.  This sum converges for $|z|<1$, and the value of
$c-a$ is obtained by dropping the pole terms and keeping only the
finite term when $z\to1$.

One difficulty with this regulator is how to extend the notion of a
KK level $p$ to the case of a generic Sasaki-Einstein
compactification.  For $S^5$ and its orbifolds, one can take $p$ to
be the usual KK level on the round $S^5$ before projection. However,
for a space like $T^{1,1}$, there is no unambiguous notion of a KK
level. Nevertheless, we have proposed a working definition of `KK
level' based on associating the $E_0$ values of the shortened
spectrum with $p$ values corresponding to what they would have been
had they come from compactification on $S^5$. Although this yields
non-integer levels $p$ for $T^{1,1}$ and its odd orbifolds, the
agreement we have found in the $c-a$ values suggests that this is a
valid regulator. In fact, the number $p$ has a clear AdS (or
CFT) interpretation for the multiplets of $S^5$
compactification: it is the number of oscillator pairs that make up
the representations of the isometry group $\mathrm{SO}(4,2) \sim \mathrm{SU}(2,2)$
\cite{Gunaydin:1984fk,Gunaydin:1998} (see also
\cite{Gunaydin:1985tc}).  It seems likely that the `KK level' we have
assigned to the multiplets on $T^{1,1}$ has a similar purely
AdS interpretation.  It would be interesting to establish
this explicitly.

Curiously, we have found that the regulated sum contributing to
$c-a$ is finite at $z=1$ for the odd orbifolds of $S^5$.  As we have
shown in Ref.~\cite{Ardehali:2013gra}, a zeta-function
regularization yields the same result as the $z^p$ regulator for
$S^5/\mathbb Z_3$; we have also checked that this is the case for
$S^5/\mathbb Z_5$, and expect it to hold for all the odd orbifolds.
However, the regulated sum does have double and single poles at
$z=1$ for the case of even orbifolds of $S^5$ and for all orbifolds
of $T^{1,1}$.  In these cases, it appears that a zeta function
regularization will produce a different result. This is something we
do not fully understand.

In fact, any time pole terms are present in the regulated $c-a$, it is
possible to shift the finite part simply by transforming $z$.  For example,
if we took the result (\ref{eq:T11ans}) for $T^{1,1}$ and let $z\to z^2$,
we would end up with $5/36+\delta_{\rm alt.~quant.}$ instead.  Of
course, such a transformation corresponds to a redefinition of the
effective KK level.  Hence this ambiguity in the finite term is closely
related to how we define the KK level.

Curiously, whenever we have found a divergent expression for $c-a$,
it has taken the form
\begin{equation}
c-a=\fft\alpha{(z-1)^2}+\fft\alpha{z-1}+\mbox{finite}=\alpha\fft{z}{(z-1)^2}+\mbox{finite}.
\end{equation}
(The third and fourth order poles that could be present always seem to
vanish when the contributions from the different multiplets are combined.)
This suggests that the pole terms may be attributed to the sum over the KK
tower as follows
\begin{equation}
\alpha\fft{z}{(z-1)^2}=\sum_{p=1}^\infty z^p(\alpha p).
\end{equation}
It would be curious to see if there is any physical interpretation of this sum
and in particular of the value of $\alpha$.  Note that $\alpha=0$ for odd
orbifolds of $S^5$, $\alpha=-3/4n$ for even orbifolds of $S^5$ and
$\alpha=-2/9n$ for orbifolds of $T^{1,1}$.

One possible way around the possible ambiguities in choosing a regulator
would be to work directly in ten-dimensional IIB supergravity.  Then instead
of summing over the KK tower and regulating this sum by multiplying by $z^p$,
we could simply use a ten-dimensional heat kernel regularization (or
nine-dimensional heat kernel for the directions transverse to the radial direction).
This can be facilitated by taking advantage of the factorizability of the heat kernel
on product spaces.  Of course, this would just replace the spectral analysis on
the internal manifold with an essentially equivalent heat kernel computation.
However, it would naturally provide a uniform regularization instead of having
separate ones involving the four-dimensional heat kernel coefficient along with
the $z^p$ regulator.

Up to a possible ambiguity due to the unknown factor $\delta_{\rm alt.~quant.}$
for the case of $T^{1,1}$, we have shown that the holographic computation of
$c-a$ in the IIB supergravity theory reproduces the corresponding field theory
result.  In particular, this leaves no room for contributions from massive string
states running in the loop.  This does not present a difficulty for the orbifolds of
$S^5$, as the string loop contribution to $c-a$ vanishes in this case \cite{Liu:2010gz}.
However, the contribution does not appear to vanish for orbifolds of $T^{1,1}$,
and adding this contribution to the supergravity result would then destroy the
perfect agreement with the dual gauge theory.  One possible explanation for
this disparity is that the string loop computation may not be completely
independent of the supergravity computation.  Although the supergravity computation
necessarily excludes massive string states, there may be overlap in
the massless sector%
\footnote{One may suspect that the massive string states would fall into long
representations of $\mathrm{SU}(2,2|1),$ much like the $\mathrm{SU}(2,2|4)$ case as discussed
in \cite{Bianchi:2003wx}, and hence would not contribute to $c-a$.  If this were the case, then the
computation of \cite{Liu:2010gz} would indeed represent a contribution from the massless sector.}.
In this case, adding the string loop result to the supergravity
result would then end up double counting some of the contributions to $c-a$.

In the course of the present work, we have received insightful
suggestions%
\footnote{We are particularly thankful to D.~Minic and L.~A.~Pando
Zayas for stimulating discussions on this point.}
as to how the $z^p$ regulator may be generalized to cases
where the KK level $p$ may be ill-defined.  A potentially
fruitful idea is to introduce separate chemical potentials
for the individual quantum numbers
(or charges) associated to the isometry group of the internal
manifold. This idea is particularly appealing when recalling the
index-like nature of the holographic $c-a$.
For example, the KK multiplets on $T^{1,1}$
are labeled by three quantum numbers $j$, $l$, $r$, corresponding
to $\mathrm{SU}(2)_j\times\mathrm{SU}(2)_l\times\mathrm U(1)_r$.
In this case, one would regulate the $T^{1,1}$ tower by multiplying
by $z_1^jz_2^lz_3^r$.  However, $j$, $l$ and $r$ are all related
to each other in the shortened towers, so it is not clear if anything
is gained by introducing separate chemical potentials for all
three quantum numbers.

Another possibility is to regulate the sum by $z^{L}$, associating a
chemical potential to the length $L$ of the superfield dual to a
given bulk multiplet.  For $S^5$ and its orbifolds, this matches the
$z^p$ regularization in the untwisted sector.  However, the length
of the dual superfield and the assigned `KK level' no longer
coincide for $T^{1,1}$ and its orbifolds. Using the $z^L$ regulator
in these cases would yield results in disagreement with the field
theory expectation (whether the massive string loop corrections
suggested in \cite{Liu:2010gz} are included or not).

Some of these regulator issues, as well as puzzles about the
possible contribution from massive string loops could potentially
be resolved by studying additional pairs of AdS/CFT duals.  A natural
extension would be to consider the Sasaki-Einstein manifolds $Y^{p,q}$.
Although knowledge of the full spectrum appears to be out
of reach, we would only need information about the shortened spectrum
in order to investigate $c-a$.  A partial analysis for $Y^{p,q}$ was performed in
\cite{Kihara:2005nt}, and we anticipate that this can be extended to provide
information on the complete shortened spectrum.  This is currently under
investigation.

\acknowledgments

We are thankful to D.~Minic and L.~A.~Pando Zayas for insightful
comments on the regularization of the sum over the KK towers.
A.~A.~A. is also grateful to G.~Dall'Agata and A.~Hanany for helpful
correspondence on the KK spectroscopy of $T^{1,1}$ and
$T^{1,1}/\mathbb{Z}_{2}$. We would also like to thank M.~Duff,
R.~Eager, Z.~Komargodski, J.~Schmude, Y.~Tachikawa and B.~Wecht for
useful correspondence. This work was supported in part by the US
Department of Energy under grants DE-SC0007859 and DE-SC0007984.

\appendix

\section{The twisted sector states for the $S^5/\mathbb Z_2$ orbifold}
\label{app:twisted}

The $S^5/\mathbb Z_2$ orbifold preserves 16 real supercharges, and
the twisted sector may be described by a six-dimensional (2,0)
theory with a single tensor multiplet \cite{Douglas:1996sw}.  The
field content is $(B_{\mu\nu}^-,5\phi,4\chi)$ transforming as the
$\mathbf1+\mathbf5+\mathbf4$ of USp(4).  This may be reduced on
AdS$_5\times S^1$ to give an effective five-dimensional $\mathcal
N=4$ spectrum classified by $\mathrm{SU}(2,2|2)$
\cite{Gukov:1998kk,Gadde:2009dj}. Making use of the decompositions
$\mathbf5\to\mathbf3_0+\mathbf1_2+\mathbf1_{-2}$ and
$\mathbf4\to\mathbf2_1+\mathbf2_{-1}$ under
$\mathrm{USp}(4)\supset\mathrm{SU}(2)_R \times\mathrm U(1)_R$, the
zero mode on the circle gives rise to the shortened $\mathcal N=4$
multiplet \cite{Gadde:2009dj}
\begin{eqnarray}
\mathcal D(2,0,0;\mathbf3_0)&=&D(2,0,0)\mathbf3_0+D(\ft52,\ft12,0)\mathbf2_{-1}
+D(\ft52,0,\ft12)\mathbf2_1+D(3,\ft12,\ft12)\mathbf1_0\nn\\
&&+D(3,0,0)\mathbf1_{-2}+D(3,0,0)\mathbf1_2,
\end{eqnarray}
where the AdS$_5$ representations are labeled by $D(E_0,s_1,s_2)$, and the
$\mathrm{SU}(2)_R\times\mathrm U(1)_R$ quantum numbers are appended.  The
non-zero-modes are also shortened.  For positive KK level $p\ge1$, we have
\cite{Gukov:1998kk,Gadde:2009dj}
\begin{eqnarray}
\mathcal D(p+1,0,0;\mathbf1_{2p+2})&=&D(p+1,0,0)\mathbf1_{2p+2}
+D(p+\ft32,\ft12,0)\mathbf2_{2p+1}+D(p+2,1,0)\mathbf1_{2p}\nn\\
&&+D(p+2,0,0)\mathbf3_{2p}
+D(p+\ft52,\ft12,0)\mathbf2_{2p-1}+D(p+3,0,0)\mathbf1_{2p-2}.\kern3em
\end{eqnarray}
The negative KK modes are just the conjugates of the positive ones.

The reduction of the $\mathcal N=4$ representations to $\mathcal N=2$
follows from the decomposition $\mathrm{SU}(2)_R\times\mathrm U(1)_R\supset
\mathrm U(1)_q\times\mathrm U(1)_r$, where
\begin{equation}
q=R-2T^3,\qquad r=\ft13(R+4T^3).
\end{equation}
Here $T^3$ is the Cartan generator of SU(2)$_R$.
The U(1)$_q$ normalization is chosen to match that of the untwisted sector, while
U(1)$_r$ takes the conventional normalization for the $\mathcal N=2$ $R$-charge.
The zero mode then breaks up into three $\mathcal N=2$ multiplets
\begin{equation}
\mathcal D(2,0,0;\mathbf3_0)=\mathcal D(2,0,0;-\ft43)_2+\mathcal D(2,0,0;0)_0
+\mathcal D(2,0,0;\ft43)_{-2},
\end{equation}
where the $q$-charge is subscripted.  The positive KK tower breaks up according to
\begin{eqnarray}
\mathcal D(p+1,0,0;\mathbf1_{2p+2})&=&\mathcal D(p+1,0,0;\ft23(p+1))_{2p+2}+
\mathcal D(p+\ft32,\ft12,0;\ft23(p+\ft32))_{2p}\nn\\
&&+\mathcal D(p+2,0,0;\ft23(p+2))_{2p-2}.
\end{eqnarray}
These are all shortened $\mathcal N=2$ states.  This information is
presented in Table~\ref{tbl:twisted}, where it is noted that they
all transform as singlets under the SU(2) corresponding to rotations
in the first two complex planes acted upon by the $\mathbb Z_2$
generator (\ref{eq:Z2act}).



\begin{thebibliography}{99}

\bibitem{Anselmi:1997am}
D.~Anselmi, D.~Z.~Freedman, M.~T.~Grisaru and A.~A.~Johansen,
{\sl Nonperturbative Formulas for Central Functions of Supersymmetric Gauge Theories},
Nucl.\ Phys.\ B {\bf 526} (1998) 543 [hep-th/9708042].

\bibitem{Henningson:1998gx}
M.~Henningson and K.~Skenderis,
{\sl The Holographic Weyl Anomaly},
JHEP {\bf 9807} (1998) 023 [hep-th/9806087].

\bibitem{Liu:2010gz}
J.~T.~Liu and R.~Minasian,
{\sl Computing $1/N^2$ Corrections in AdS/CFT},
arXiv:1010.6074 [hep-th].

\bibitem{Bilal:1999ph}
A.~Bilal and C.-S.~Chu,
{\sl A Note on the Chiral Anomaly in the AdS / CFT Correspondence and $1 / N^2$ Correction},
Nucl.\ Phys.\ B {\bf 562} (1999) 181 [hep-th/9907106].

\bibitem{Bilal:1999ty}
A.~Bilal and C.-S.~Chu,
{\sl Testing the AdS / CFT Correspondence Beyond Large $N$},
hep-th/0003129.

\bibitem{Mansfield:2000zw}
P.~Mansfield and D.~Nolland,
{\sl Order $1 / N^2$ Test of the Maldacena Conjecture: Cancellation of the One Loop
Weyl Anomaly},
Phys.\ Lett.\ B {\bf 495} (2000) 435 [hep-th/0005224].

\bibitem{Mansfield:2002pa}
P.~Mansfield, D.~Nolland and T.~Ueno,
{\sl Order $1 / N^2$ Test of the Maldacena Conjecture. 2. the Full Bulk One Loop
Contribution to the Boundary Weyl Anomaly},
Phys.\ Lett.\ B {\bf 565} (2003) 207 [hep-th/0208135].

\bibitem{Mansfield:2003gs}
P.~Mansfield, D.~Nolland and T.~Ueno,
{\sl The Boundary Weyl Anomaly in the ${\mathcal{N}}\!=4$ SYM / Type IIB
Supergravity Correspondence},
JHEP {\bf 0401} (2004) 013 [hep-th/0311021].

\bibitem{Gunaydin:1984fk}
M.~Gunaydin and N.~Marcus,
{\sl The spectrum of the $S^5$ compactification of the chiral $N=2$, $D=10$ supergravity
and the unitary supermultiplets of $U(2, 2/4)$},
Class.\ Quant.\ Grav.\  {\bf 2}, L11 (1985).

\bibitem{Kim:1985ez}
H.~J.~Kim, L.~J.~Romans and P.~van Nieuwenhuizen,
{\sl Mass spectrum of chiral ten-dimensional $N=2$ supergravity on $S^5$},
Phys.\ Rev.\ D {\bf 32}, 389 (1985).

\bibitem{Gauntlett:2004yd}
J.~P.~Gauntlett, D.~Martelli, J.~Sparks and D.~Waldram,
{\sl Sasaki-Einstein metrics on $S^2\times S^3$},
Adv.\ Theor.\ Math.\ Phys.\  {\bf 8}, 711 (2004) [hep-th/0403002].

\bibitem{Cvetic:2005ft}
M.~Cvetic, H.~Lu, D.~N.~Page and C.~N.~Pope,
{\sl New Einstein-Sasaki spaces in five and higher dimensions},
Phys.\ Rev.\ Lett.\  {\bf 95}, 071101 (2005) [hep-th/0504225].

\bibitem{Ardehali:2013gra}
A.~A.~Ardehali, J.~T.~Liu and P.~Szepietowski,
{\sl The Spectrum of IIB supergravity on $\mathrm{AdS}_5\times S^5/\mathbb Z_3$
and a $1/N^2$ test of AdS/CFT},
JHEP {\bf 1306} (2013) 024 [arXiv:1304.1540 [hep-th]].

\bibitem{Anselmi:1998zb}
D.~Anselmi and A.~Kehagias,
{\sl Subleading Corrections and Central Charges in the AdS / CFT Correspondence},
Phys.\ Lett.\ B {\bf 455} (1999) 155 [hep-th/9812092].

\bibitem{Christensen:1978gi}
S.~M.~Christensen and M.~J.~Duff,
{\sl Axial and Conformal Anomalies for Arbitrary Spin in Gravity and Supergravity},
Phys.\ Lett.\ B {\bf 76} (1978) 571.

\bibitem{Freedman:1999gp}
D.~Z.~Freedman, S.~S.~Gubser, K.~Pilch and N.~P.~Warner, {\sl
Renormalization Group Flows from Holography--Supersymmetry and a
$c$-Theorem}, Adv.\ Theor.\ Math.\ Phys. {\bf 3} (1999) 363
[arXiv:hep-th/9904017].

\bibitem{Gadde:2011}
A.~Gadde, L.~Rastelli, S.~S.~Razamat, and W.~Yan, {\sl On the
superconformal index of $\mathcal N = 1$ IR fixed points. A holographic check},
JHEP {\bf 1103} (2011) 1 [arXiv:1011.5278 [hep-th]].

\bibitem{Gukov:1998kk}
S.~Gukov,
{\sl Comments on ${\mathcal{N}}\!=2$ AdS orbifolds},
Phys.\ Lett.\ B {\bf 439}, 23 (1998) [hep-th/9806180].

\bibitem{Gadde:2009dj}
A.~Gadde, E.~Pomoni and L.~Rastelli,
{\sl The Veneziano Limit of ${\mathcal{N}}\!=2$ Superconformal QCD:
Towards the String Dual of ${\mathcal{N}}\!=2$ $SU(N_c)$ SYM with
$N_f = 2 N_c$},
arXiv:0912.4918 [hep-th].

\bibitem{Ceresole:1999zs}
A.~Ceresole, G.~Dall'Agata, R.~D'Auria and S.~Ferrara,
\emph{Spectrum of type IIB supergravity on AdS$_5\times T^{11}$:
Predictions on $\mathcal N=1$ SCFT's},
Phys.\ Rev.\ D {\bf 61}, 066001 (2000) [hep-th/9905226].

\bibitem{Ceresole:1999ht}
A.~Ceresole, G.~Dall'Agata and R.~D'Auria,
\emph{KK spectroscopy of type IIB supergravity on AdS$_5\times T^{11}$},
JHEP {\bf 9911}, 009 (1999) [hep-th/9907216].

\bibitem{Eager:2012hx}
R.~Eager, J.~Schmude and Y.~Tachikawa,
{\sl Superconformal Indices, Sasaki-Einstein Manifolds, and Cyclic Homologies},
arXiv:1207.0573 [hep-th].

\bibitem{Cassani:2010uw}
D.~Cassani, G.~Dall'Agata and A.~F.~Faedo, {\sl Type IIB
Supergravity on Squashed Sasaki-Einstein Manifolds}, JHEP {\bf 1005}
(2010) 094 [arXiv:1003.4283 [hep-th]].

\bibitem{Liu:2010sa}
J.~T.~Liu, P.~Szepietowski and Z.~Zhao, {\sl Consistent Massive
Truncations of IIB Supergravity on Sasaki-Einstein Manifolds},
Phys.\ Rev.\ D {\bf 81} (2010) 124028 [arXiv:1003.5374 [hep-th]].

\bibitem{Gauntlett:2010vu}
J.~P.~Gauntlett and O.~Varela, {\sl Universal Kaluza-Klein
Reductions of Type IIB to ${\mathcal{N}}\!=4$ Supergravity in Five
Dimensions}, JHEP {\bf 1006} (2010) 081 [arXiv:1003.5642 [hep-th]].

\bibitem{Skenderis:2010vz}
K.~Skenderis, M.~Taylor and D.~Tsimpis, {\sl A Consistent Truncation
of IIB Supergravity on Manifolds Admitting a Sasaki-Einstein
Structure}, JHEP {\bf 1006} (2010) 025 [arXiv:1003.5657 [hep-th]].

\bibitem{Nolland:2003kc}
D.~Nolland,
{\sl AdS / CFT Boundary Conditions, Multitrace Perturbations, and the $c$-Theorem},
Phys.\ Lett.\ B {\bf 584} (2004) 192 [hep-th/0310169].

\bibitem{Gubser:2002zh}
S.~S.~Gubser and I.~Mitra,
{\sl Double trace operators and one-loop vacuum energy in AdS / CFT},
Phys.\ Rev.\ D {\bf 67} (2003) 064018 [hep-th/0210093].

\bibitem{Gubser:2002vv}
S.~S.~Gubser and I.~R.~Klebanov,
{\sl A universal result on central charges in the presence of double-trace deformations},
Nucl.\ Phys.\ B {\bf 656} (2003) 23 [hep-th/0212138].

\bibitem{Hartman:2006dy}
T.~Hartman and L.~Rastelli,
{\sl Double-Trace Deformations, Mixed Boundary Conditions and Functional
Determinants in AdS/CFT},
JHEP {\bf 0801} (2008) 019 [hep-th/0602106].

\bibitem{Franco:2005fd}
S.~Franco, A.~Hanany and A.~M.~Uranga,
{\sl Multi-Flux Warped Throats and Cascading Gauge Theories},
JHEP {\bf 0509} (2005) 028 [hep-th/0502113].

\bibitem{Gunaydin:1998}
M.~Gunaydin, D.~Minic and M.~Zagermann, {\sl 4D Doubleton Conformal
Theories, CPT and IIB String on AdS$_5\times S^5$}, Nucl.\
Phys.\ B {\bf 534} (1998) 96 [hep-th/9806042].

\bibitem{Gunaydin:1985tc}
M.~G\"unaydin and N.~P.~Warner,
{\sl Unitary Supermultiplets of Osp(8/4,$\mathbb R$) and the Spectrum of the $S^7$
Compactification of Eleven-Dimensional Supergravity},
Nucl.\ Phys.\ B {\bf 272} (1986) 99.

\bibitem{Bianchi:2003wx}
M.~Bianchi, J.~F.~Morales and H.~Samtleben, {\sl On stringy AdS$_5\times
S^5$ and higher spin holography}, JHEP {\bf 0307}, 062 (2003)
[hep-th/0305052].

\bibitem{Kihara:2005nt}
H.~Kihara, M.~Sakaguchi and Y.~Yasui,
{\sl Scalar Laplacian on Sasaki-Einstein Manifolds $Y^{p,q}$},
Phys.\ Lett.\ B {\bf 621} (2005) 288 [hep-th/0505259].

\bibitem{Douglas:1996sw}
M.~R.~Douglas and G.~W.~Moore,
{\sl D-Branes, Quivers, and ALE Instantons},
hep-th/9603167.

\end{thebibliography}
\end{document}